\documentclass[twocolumn,showpacs,preprintnumbers,amsmath,amssymb]{revtex4}

\usepackage{graphics}
\usepackage{graphicx}
\usepackage{dcolumn}
\usepackage{bm}
\usepackage{amssymb}
\begin{document}
\preprint{APS/123-QED}

\title{Magneto-elastic coupling in hexagonal YMnO$_3$}

\author{Mario  Poirier}
\author{Francis Lalibert\'e}
\affiliation{ Regroupement Qu\'eb\'ecois sur les Mat\'eriaux de
Pointe, D\'epartement de Physique, Universit\'e de Sherbrooke,
Sherbrooke, Qu\'ebec,Canada J1K 2R1}
\author{Loreynne Pinsard}
\author{Alexandre Revcolevschi}
\affiliation{Laboratoire de Chimie des Solides, Universit\'e
Paris-Sud, 91405 Orsay C\'edex, France}

\date{\today}

\begin{abstract}

We report an ultrasonic investigation of the elastic moduli on a
single crystal of hexagonal YMnO$_3$ as a function of
temperature. Stiffening anomalies in the antiferromagnetic N\'eel
state below $T_N$ = 72.4 K are observed on all the four elastic
moduli $C_{ii}$. The anomalies are the most important on $C_{11}$
and $C_{66}$ for in-plane elastic deformations; this is
consistent with a strong coupling of the lattice with the
in-plane exchange interactions. We use a Landau free energy model
to account for these elastic anomalies. We derive an expression
which relates the temperature profile of the anomaly to the order
parameter; the critical exponent associated to this parameter
$\beta$ = 0.42 is not consistent with a chiral XY or 3D
Heisenberg universality class, but more in agreement with a
conventional antiferromagnetic long range order. A tiny softening
anomaly on $C_{11}$ for which hysteresis effects are observed
could be indicative of an interaction between ferroelectric and
magnetic domains at $T_N$. Moreover, magnetic fluctuations
effects both above and below $T_N$ are identified through
abnormal temperature and magnetic field effects.

\end{abstract}

\maketitle
\section{INTRODUCTION}
Hexagonal YMnO$_3$ is one of the most well-known examples of
multiferroics, a material that possesses both ferroelectricity
and magnetism. The ferroelectric (FE) transition occurs at a
relatively high temperature above 900 K. Below the Curie
temperature the compound crystallizes in a hexagonal structure
with the space group $P6_3cm$ \cite{YAKEL1963}. The Mn$^{3+}$
($S$=2) ions form nearly triangular networks in $z$=0 and $z$=1/2
layers which stack in the ABAB sequence along the $z$ axis with a
wide separation introduced by the intervening Y and O ions
\cite{KATSUFUJI2002}. This suggests a predominant two-dimensional
(2D) character in the $ab$ plane. For this system the geometrical
frustration of the antiferromagnetic (AF) spins on the triangular
lattice results in the $120^{\rm o}$ spin ordering below the
N\'eel temperature $T_N$ $\sim$ 70 K.

The Curie-Weiss (CW) temperature deduced from susceptibility
measurements is approximately ten times higher than $T_N$
\cite{KATSUFUJI2002,MUNOZ2000,KATSUFUJI2001} and the magnetic
moment is lower than the expected value from the fully polarized
Mn$^{3+}$ \cite{KATSUFUJI2002,MUNOZ2000,BIERINGER1999}. This
reduction in $T_N$ and magnetic moment is consistent with strong
spin fluctuations due to geometrical frustration and/or low
dimensionality. The coexistence of novel magnetism and
ferroelectricity suggests a peculiar interaction between the
spins and the lattice degrees of freedom, although the low $T_N$,
compared to the ferroelectric transition, infers weak coupling
between magnetism and ferroelectricity. Although various
crystallographic, optical and magnetic  structural studies have
been performed to characterize independently the FE and magnetic
properties \cite{YAKEL1963,ILIEV1997,MUNOZ2000}, only a few ones
have been directed to understand their interrelation. Dielectric
constant \cite{KATSUFUJI2001}, specific heat
\cite{TACHIBANA2005,TOMUTA2001} and thermal conductivity
\cite{SHARMA2004,ZHOU2006} experiments revealed a spin-lattice
coupling whose understanding remains an open question. As far as
magnetism is concerned, neutron scattering studies
\cite{LONKAI2002,SATO2003,PARK2003,ROESSLI2005} have reported the
presence of unconventional spin fluctuations both above and below
$T_N$ that could be related to the anomalous increase of the
thermal conductivity upon magnetic ordering \cite{SHARMA2004}.
However, the question of coexistence of long-range order with
significant spin-fluctuations in the N\'eel state for the
triangular lattice remains. The symmetry of the order parameter
and the nature of the universality class associated with the
$120^{\rm o}$ spin structure are also not resolved up to now.

To investigate the coupling between magnetism and the lattice
degrees of freedom in YMnO$_3$, we report an ultrasonic
investigation of the elastic moduli of a single crystal in the
temperature range 2-150 K, both above and below the spin ordering
temperature $T_N$. Although elastic anomalies at $T_N$ are
observed on all studied moduli, an important coupling of the
magnetic order parameter is only obtained with an in-plane strain
field. A Landau free energy expansion is used to interpret these
anomalies and a critical analysis of the temperature behavior
just below $T_N$ is tentatively given. Small temperature
hysteretic effects are observed indicating a domain structure. A
peculiar temperature profile above $T_N$ for two moduli is
discussed in relation with the magnetic fluctuation issue.

\section{EXPERIMENT}
High quality YMnO$_3$ single crystals are grown using the standard
floating zone technique. Powder X-ray diffraction measurements
confirmed that the crystal investigated here has the $P6_3cm$
hexagonal structure at room temperature ($a$=$b$=6.1380(1)\AA ,
$c$=11.4045(1)\AA). A single crystal was then oriented with a
back Laue X-ray diffraction technique and cut with a wire-saw to
obtain two sets of parallel faces respectively oriented
perpendicular (separation distance 3.29 mm) and parallel (3.54
mm) to the hexagonal $c$ axis. The faces were subsequently
polished to obtain a mirror like aspect. We used a pulsed-echo
ultrasonic technique in the reflection mode to measure the
velocity and amplitude of longitudinal and transverses waves
propagating along both directions. The plane acoustic waves were
generated with LiNbO$_3$ piezoelectric transducers (fundamental
frequency 30 MHz and odd overtones) bonded to one parallel face
with silicon seal. The technique consists in measuring the phase
shift and the amplitude of the first elastic pulse reflected from
the opposite parallel face. Doing so, this technique yields only
velocity and attenuation variations; an absolute value of the
velocity is obtained by measuring the transit time between
different reflected echoes. The crystal was mounted in a Variable
Temperature Insert (VTI); the useful temperature range was 2-150
K, the maximum value being imposed by phase transformation of the
silicon seal around 200 K. A magnetic field up to 14 Tesla could
be applied along different directions.

For a crystal having hexagonal symmetry the elastic modulus matrix
has only five non-zero independent components $C_{ij}$
\cite{DIEULESAINT}. Four of these components can be obtained by
propagating ultrasonic waves parallel and perpendicular to the
hexagonal $c$ axis and by measuring the velocity of these waves
with various polarizations (longitudinal and transverse). For
propagation along $c$, the longitudinal and transverse velocities
are respectively given by $V_{L1} = ({C_{33}/\rho})^{1 \over 2}$
and $V_{T1} = ({C_{44}/\rho})^{1 \over 2}$; for the perpendicular
direction, besides the longitudinal one, two transverse
velocities (polarization $\perp c$ and $\parallel c$) are
obtained: $V_{L2} = ({C_{11}/\rho})^{1 \over 2}$, $V_{T\perp} =
({C_{66}/{\rho}})^{1 \over 2}$ and $V_{T\parallel} =
({C_{44}/{\rho}}) ^{1 \over 2}$. By using the density $\rho = 5.1$
g/cm$^3$ and the experimental velocities, the four $C_{ij}$ are
obtained. We notice that $V_{T1}$ and $V_{T\parallel}$ are equal
if the crystal is correctly oriented, a situation that was
checked in our experiment. The absolute values of the moduli at
low temperatures are given in Table I.

\begin{table}[h]
\caption{\label{table1}Elastic moduli of YMnO$_3$ (units $10^{10}
N/m^2$) at 4K.}

\begin{ruledtabular}
\begin{tabular}{cccc}
$C_{11}$ &   $C_{33}$   &   $C_{44}$   & $C_{66}$
\\ \hline
& & & \\
$18.5 \pm 0.1$ & $29.8 \pm 0.5$ & $9.86 \pm 0.15$ & $5.94 \pm
0.05$
\\
\end{tabular}
\end{ruledtabular}
\end{table}

\section{RESULTS}
The coupling between the lattice and the spin degrees of freedom
generally yields anomalies on selected elastic moduli when
three-dimensional magnetic order sets in. The selection is imposed
by the nature of the coupling between the elastic deformation and
the order parameter, coupling which is highly sensitive on the
symmetry of the latter. This coupling is confirmed for all elastic
moduli investigated in the present work on multiferroic hexagonal
YMnO$_3$. Our ultrasonic velocity measurements have been
performed at different frequencies between 30 and 300 MHz; no
significative frequency dependence were noticed on the velocity.
The data presented below were all obtained at 103 MHz.

\begin{figure}[H,h]
\includegraphics[width=8.5cm]{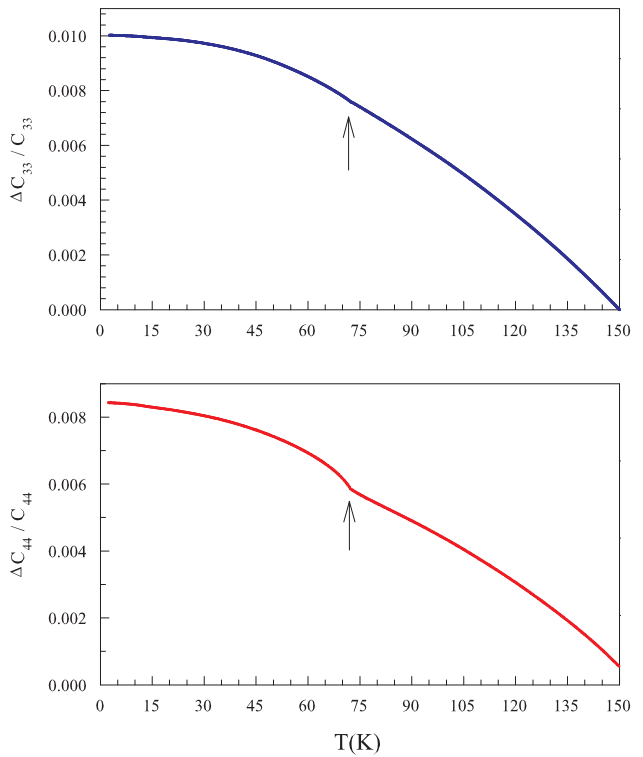} \caption{\label{fig:fig1}
Temperature profile of $\Delta C_{33}/C_{33}$ and $\Delta
C_{44}/C_{44}$. $T_N$ is indicated by an arrow.}
\end{figure}

In figure 1 we present the temperature profile of the relative
variation of the two elastic moduli that are the least affected by
magnetic ordering, $\Delta C_{33}/C_{33}$ and $\Delta
C_{44}/C_{44}$. When the temperature is decreased from 150 K, the
modulus $C_{33}$ which implies a pure compression along the $c$
axis smoothly increases and saturates at low temperature. This
temperature profile appears normal in view of the expected
stiffening following the progressive disappearance of phonon
anharmonic effects at low temperatures. There is however a small
increase of slope below 72 K indicating a further stiffening when
AF magnetic ordering occurs at $T_N$. C$_{44}$, related to a shear
deformation in a plane containing the $c$ axis, presents a
similar temperature profile. There is however a much more
pronounced upturn of the modulus below 72.5 K confirming a
definite stiffening below $T_N$ ($\sim 0.1\%$).

\begin{figure}[H,h]
\includegraphics[width=8.5cm]{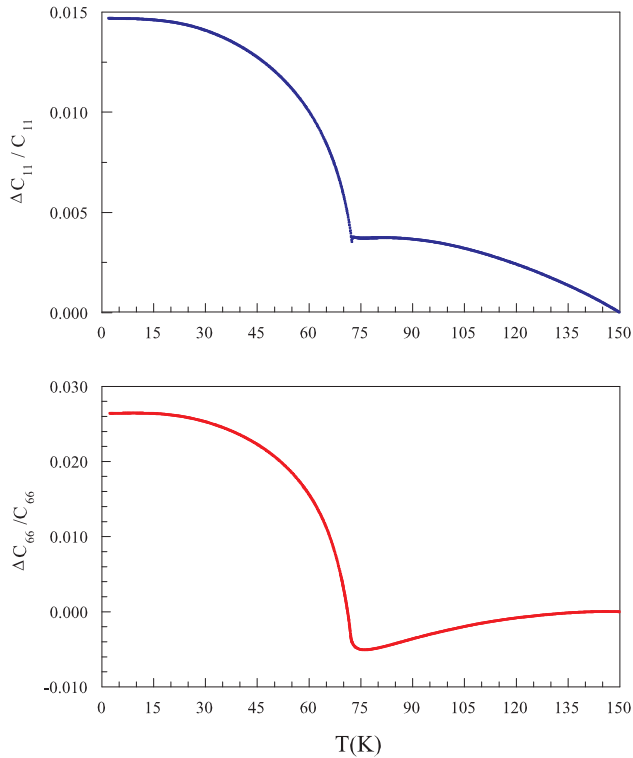} \caption{\label{fig:fig2} Temperature profile of $\Delta C_{11}/C_{11}$ and $\Delta
C_{66}/C_{6}$.}
\end{figure}

The temperature profile of the relative variation of the other two
moduli appears both similar and different from the previous ones
and, in a sense, less normal. These data are presented in figure
2 over the same temperature range. $C_{11}$,  corresponding to a
pure compression in the $ab$ plane, increases weakly below 150 K
and, on this scale, appears to saturate around 90 K. Then, an
abrupt stiffening begins at $T_N$ = 72.4 K until saturation is
obtained below 15 K; the amplitude of the stiffening is
approximatively one percent. $C_{66}$ is also associated to
in-plane deformations and it shows a non conventional behavior.
The modulus decreases below 150 K until a minimum is reached
around 75 K; then it increases further ($\sim 3\%$) with a clear
acceleration near 72 K. Below 72 K the thermal profile resembles
the one observed on $C_{11}$.

It appears then that the four elastic moduli investigated here,
$C_{11}$, $C_{33}$, $C_{44}$ and $C_{66}$ stiffen progressively
below the N\'eel temperature $T_N$ = 72.4 K. Moreover, the
temperature profile of the anomalies seems to mimic the AF
magnetic order parameter, the magnetic moment $\textbf m(T)$
\cite{MUNOZ2000,LONKAI2002}. It is worth to notice that
significant elastic anomalies are observed for moduli implying
elastic deformation (compression and shear) only in the $ab$
plane, $C_{11}$ and $C_{66}$; these are also the only moduli to
present an unconventional elastic behavior above $T_N$. These
observations could appear consistent with a two-dimensional
character of the magnetic properties. Moreover, since the anomaly
on $C_{33}$ is very small, this means that inter-plane exchange
interactions $J_{1}^{'}$ and $J_{2}^{'}$ are not strongly coupled
to the lattice compared to in-plane ones $J_1$ and $J_2$ which
yield an important anomaly on $C_{11}$ in agreement with the
analysis of spin fluctuations \cite{SATO2003}. The analysis of the
temperature dependence of these anomalies is highly delicate
since it necessitates a precise knowledge of the overall
temperature background both above and below $T_N$. Such an
analysis appears difficult for $C_{33}$ because the amplitude of
the anomaly is too small compared to the total modulus variation
with temperature. However this could be done for $C_{11}$ and
$C_{66}$ since the amplitude of the anomalies, respectively
around {1\% and 3\%}, are much larger. It will also be tempted
for $C_{44}$ even if the variation is only around 0.1\%.

\begin{figure}[H,h]
\includegraphics[width=8.5cm]{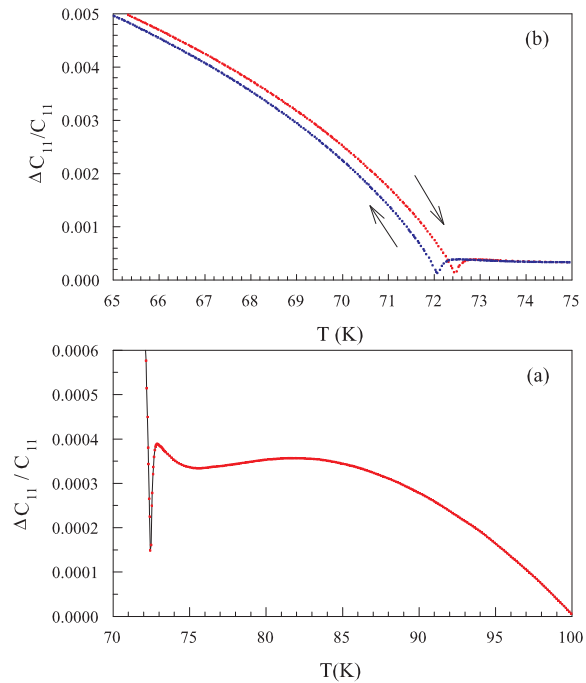}
\caption{\label{fig:fig3} Temperature profile of the relative
variation of the elastic modulus $C_{11}$: (a) reduced scales for
T $>$ $T_N$; (b) hysteresis near $T_N$, arrows indicating the
direction of the temperature sweeps.}
\end{figure}

Before doing so, let us examine more precisely the temperature
profile of $C_{11}$ just above $T_N$. The relative variation
${\Delta C_{11}}/{C_{11}}$ (relative to the value at 100 K) is
presented in figure 3a. The modulus $C_{11}$ increases smoothly
as the temperature is decreased from 100 K; then it goes through
a maximum around 82 K, reaches a minimum just above 75 K and
increases further before reaching the AF transition. A sharp
downward step ($0.02 \%$) occurs below 72.8 K before an abrupt
upturn at 72.4 K signaling a progressive stiffening. These
features in this temperature range were observed at all
frequencies without any significant modification. This abrupt
downward step preceding a stiffening is accompanied by a small
hysteresis (0.4 K) when sweeping the temperature up and down
through the transition as shown in figure 3b. The temperature
profile above $T_N$ is only modified up to 75 K where a local
minimum is observed (Fig.3a). Although they show also hysteresis,
the other moduli do not show a downward step at $T_N$. A minimum
is however observed on $C_{66}$ near 75 K (Fig.2), minimum that
could be a signature of $C_{11}$ since $C_{66} = \frac{1}{2}
(C_{11} -C_{12})$.

\begin{figure}[H,h]
\includegraphics[width=8.5cm]{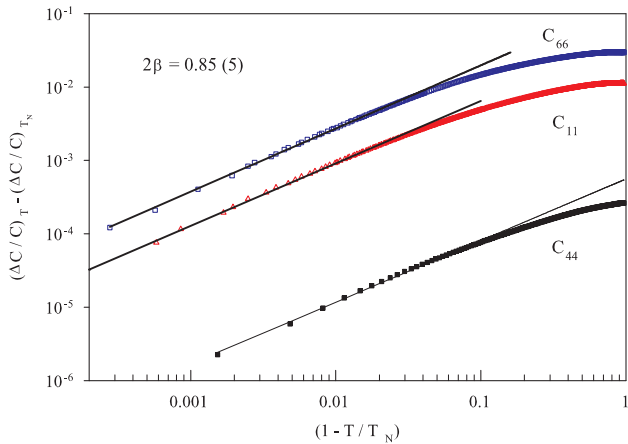} \caption{\label{fig:fig4}
Relative variation of the elastic modulus $C_{11}$, $C_{44}$ and
$C_{66}$ as a function of the reduced temperature $(1 - T /
T_N)$. Straight lines represent fits to Eq.(7).}
\end{figure}

\section{DISCUSSION}
The effects of magnetic frustration in the stacked triangular
lattice have been the object of many studies over the years
\cite{COLLINS1997} and ultrasonic techniques
\cite{TRUDEAU1993,QUIRION2005,QUIRION2006} have been used
extensively as a mean to obtain the magnetic phase diagram of
ABX$_3$ systems. Similarly to what has been done in these
systems, we can use a phenomenological Landau model to take
account of the observed changes in the elastic properties of
YMnO$_3$. If one takes into account the coupling between the
order parameter $Q$ and the strain $e_i$, three terms contribute
to the free energy $F(Q,e_i)$.
\begin{equation}
F(Q,e_i) = L(Q) + F_{el}(e_i) + F_c(Q,e_i)
\end{equation}
where $L(Q)$ is the Landau-type free energy expressed in terms of
a power series of the order parameter $Q$, $F_{el}(e_i)$ the
elastic energy associated to the deformations and $F_c(Q,e_i)$ the
coupling between the strain $e_i$ and the order parameter $Q$.
Because of the $120^{\rm o}$ spin arrangement in the $ab$ plane
\cite{MUNOZ2000,SATO2003} in the absence of a magnetic field, we
can write with in-plane spin component $S_\perp$ as the order
parameter (we neglect any $S_z$ component)
,
\begin{equation}
L(S_\perp) = aS_{\perp}^{2} + bS_{\perp}^{4}
\end{equation}
If we neglect any terms associated to the elastic moduli $C_{12}$
and $C_{13}$, the elastic energy for a hexagonal structure is
given by \cite{DIEULESAINT}
\begin{align}
F_{el}(e_i) = & \tfrac{1}{2}\, C_{11}(e_{1}^{2} + e_{2}^{2}) +
\tfrac{1}{2}\, C_{33}e_{3}^{2} + \tfrac{1}{2}\, C_{44}(e_{4}^{2}
+ e_{5}^{2}) \notag
\\
& + \tfrac{1}{2}\, C_{66}e_{6}^{2}
\end{align}
where $C_{ij}$ represent the bare elastic moduli in the high
temperature phase. For the coupling term, time inversion symmetry
imposes a quadratic dependence on the order parameter,
$S_{\perp}^{2}$: the coupling term associated to the strain $e_i$
can thus be written as
\begin{align}
F_c(S_\perp,e_i) = g_{i,r}S_{\perp}^{2}e_{i}^{r}
\end{align}
where $g_{i,r}$ is the coupling constant associated to the
deformation $e_i$ elevated to the power $r$. The term with $r =
2$ is always allowed whatever the symmetry but $r = 1$ implies
inequivalence of $| e_i|$ and $-|e_i|$. If we minimize the Landau
energy relative to the strain $e_i$, we obtain a spontaneous
strain $e_i(S_\perp)$ which is used to derive the elastic constant
$C_{ij}^{'}$ \cite{REWALD}. The biquadratic coupling term ($r =
2$) leads to a stiffening of the elastic constant $C_{ii}$
according to the general form
\begin{align}
C_{ii}^{'} = C_{ii} + 2 g_{i,2} S_{\perp}^{2}
\end{align}
For the case of linear-quadratic coupling ($r = 1$), a negative
step is obtained according to
\begin{align}
C_{ii}^{'} = C_{ii} - \frac{g_{i,1}^{2}}{2b}
\end{align}

The results presented in figures 1 and 2 are consistent with the
prediction of Eq.(5), a stiffening related to ${S_\perp}^2$. We
give in Table II the relative value of the coupling constant
$g_{i,2}$ as deduced from figures 1 and 2. The largest constant
is obtained for $C_{11}$ and  $g_{1,2}$ is thus taken as 1;
$g_{6,2}$ associated to $C_{66}$, is also very important around
0.9. The constants $g_{3,2}$ and $g_{4,2}$ are found smaller by a
factor between 20-25. This anisotropy of the coupling to the order
parameter appears consistent with the one observed in the
in-plane and inter-plane exchange constants \cite{SATO2003} and
with a XY symmetry.

\begin{table}[h]
\caption{\label{table1}Relative values of the coupling constant
$g_{i,2}$.}
\begin{ruledtabular}
\begin{tabular}{cccc}
$g_{1,2}$ & $g_{3,2}$ & $g_{4,2}$ & $g_{6,2}$ \\
\hline
&  &  &  \\
1.0 &0.04(1) & 0.05(1)  & 0.90(5)\\
\end{tabular}
\end{ruledtabular}
\end{table}

We have then tried to analyze the variation of the elastic moduli
$C_{11}$, $C_{44}$ and $C_{66}$ relative to their value at $T_N$
by fitting the order parameter to $S_\perp \sim (1 - T /
T_N)^\beta$ near the critical temperature $T_N$.
\begin{align}
{{\Delta C_{ii}}\over {C_{ii}}} = D_i (1 - T / T_N)^{2\beta}
\end{align}
where $D_i$ is a constant depending on $g_{i,2} / C_{ii}$. This
analysis for $C_{11}$ and $C_{66}$ is surely more valid than for
$C_{44}$ since the anomalies are at least 10 times larger at
$T_N$. The results of this analysis are presented in figure 4 on a
log-log scale. Not only the overall temperature dependence
appears very similar for $C_{11}$ and $C_{66}$, but the same
exponent $2\beta = 0.85(5)$ is obtained with $T_N$ = 72.43(5).
Surprisingly the same exponent is also found for $C_{44}$, what
means that the overall temperature profile of the background near
$T_N$ is not important to determine the critical behavior. Also
hysteresis effects were found to only affect $T_N$ (Fig.3), not
the exponent. The value $\beta = 0.42(3)$, which is obtained over
only two decades of the reduced temperature, is neither
consistent with chiral XY (0.25) and chiral Heisenberg (0.30)
universality classes or with XY (0.35) and Heisenberg (0.36) ones
\cite{COLLINS1997}. Although this appears to contradict our
previous observation pertaining to the coupling constants, it is
clear from Fig.4 that the same physical phenomenon is at the
origin of the elastic anomalies on $C_{11}$ and $C_{66}$, and by
extension on the other two moduli. The AF magnetic order parameter
$S_\perp$ has been associated with 3D Heisenberg or 3D XY
symmetry classes from specific heat \cite{TACHIBANA2005} and
muon-spin relaxation \cite{LANCASTER2007} studies; our study is
however not able to resolve this issue.

Finally, the sharp downward step observed on $C_{11}$ (Fig.3a)
could be consistent with the prediction of Eq.(6). On the one
hand, the observation of hysteresis in the vicinity of this peak
could be consistent with weak ferromagnetism, although this has
been observed only in ScMnO$_3$ \cite{MUNOZ2000}. On the other
hand, coupling between magnetic and electric domains in YMnO$_3$
has been reported by imaging with optical second harmonic
generation \cite{FIEBIG2002}; it has been suggested that the
small anomaly observed on the in-plane dielectric constant
\cite{HUANG1997} could result from such coupling effects on
domain walls \cite{QIU2006}. The small peak on $C_{11}$ at $T_N$
together with the observation of hysteresis could result from an
interaction between in-plane compression waves and the domain
walls.

\begin{figure}[H,h]
\includegraphics[width=8.5cm]{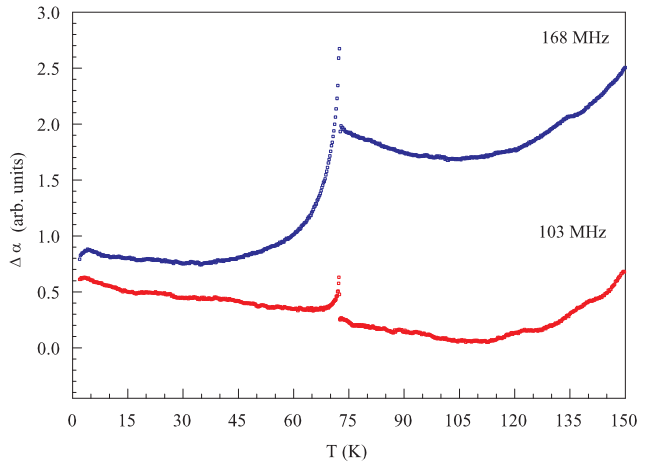} \caption{\label{fig:fig5}
Variation of the attenuation $\Delta \alpha$ of in-plane
compression waves a function of temperature at two frequencies,
103 and 168 MHz.}
\end{figure}

Previous neutron scattering studies
\cite{SATO2003,PARK2003,ROESSLI2005} reported the presence of
unconventional spin fluctuations both above and below $T_N$.
These fluctuations could explain the anomalous behavior of
$C_{11}$ (and by extension $C_{66}$) well above $T_N$ as shown in
figure 3a. Since similar effects are not observed for $C_{33}$
and $C_{44}$, fluctuations preferentially 2D-like associated to
in-plane exchange constants are likely responsible. The ultrasonic
attenuation is generally more sensitive to fluctuations since it
is a dissipative phenomenon. These attenuation variations $\Delta
\alpha$ are difficult to obtain with precision on a wide
temperature range and this is why they must be interpret with
great care. We present in figure 5 the variation of the
attenuation associated to the $C_{11}$ modulus as a function of
temperature at two frequencies. As the temperature is decreased
from high temperature, the attenuation generally decreases
because the scattering of the acoustic waves by phonons is
reduced. Here the attenuation shows a minimum at a rather high
temperature near 110 K for both frequencies and it increases for
lower temperatures, a behavior that is not consistent with
decreasing phonon scattering. At 103 MHz, the AF magnetic
ordering produces a very sharp peak at $T_N$ but does not change
significantly the monotonous increase of the attenuation that
begun near 110 K and the attenuation increases down to the lowest
temperature. At 168 MHz, the temperature profile from 150 K to 72
K is conserved; however, the peak is more important at $T_N$ and
the attenuation decreases rapidly below the N\'eel temperature.
At lower temperature, there is no saturation of the attenuation
but a small increase below 30 K is observed. These attenuation
data strongly suggest that spin degrees of freedom, in other
words magnetic fluctuations both above (at least up to 110 K) and
below $T_N$, dissipate acoustic energy. At $T_N$, the sharp
increase is due to the spontaneous strain yielding to eq.(6) and
for which hysteresis is observed. Below $T_N$, spin degrees of
freedom are lost when 3D AF ordering occur and the attenuation
decreases rapidly. The data reveal an increase of attenuation at
temperatures much higher than $T_N$ and a reduction below when 3D
AF magnetic ordering occurs. Although this reduction increases
rapidly in amplitude with frequency, dissipation still occurs
well below $T_N$. These observations are consistent with
\textit{extraordinary spin-phonon interactions} deduced from
thermal conductivity data \cite{SHARMA2004}. How these
fluctuations are related to the domain structure remains an open
question. The presence of domains and fluctuations near $T_N$
could complicate the analysis of the critical behavior presented
in Fig.4 and affect the determination of the exponent $\beta$.

\begin{figure}[H,h]
\includegraphics[width=8.5cm]{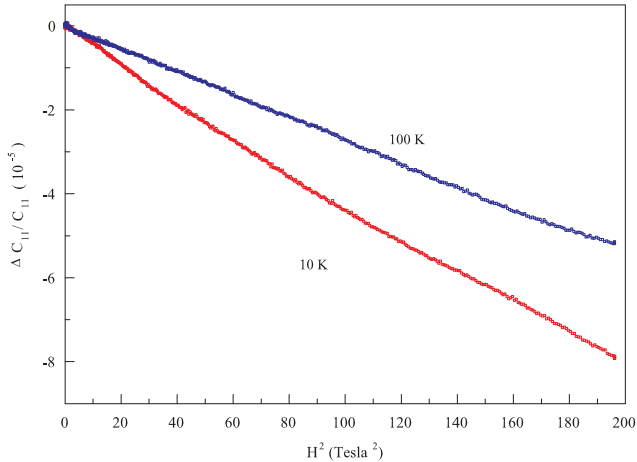} \caption{\label{fig:fig6}
Relative variation of $C_{11}$ as a function of the magnetic
field at two temperatures well below and above $T_N$.}
\end{figure}

The magnetic character of the peculiar behavior of $C_{11}$ above
$T_N$ can also be revealed by a magnetic field analysis. Up to 14
Tesla we did not find any measurable variation of $T_N$ whatever
the orientation of the field relative to the $c$ axis was. This
probably implies that the amplitude of the magnetic order
parameter below $T_N$ is highly rigid relative to the magnetic
field. Nevertheless some magnetic field effects can be measured
on $C_{11}$ over the 2-150 K temperature range. These are
presented in figure 6 as $C_{11}$ variations relative to the
value at zero field. The data are presented as a function of
$H^2$ at two temperatures, 10 K well below $T_N$ and 100 K well
above for a field applied parallel to the $c$ axis. The field
profile is very similar for both temperatures and the absolute
value has not even decreased by a factor 2 when the temperature
has been multiplied by ten. These effects have thus to be related
to the overall background: a magnetic field enhances the
softening of $C_{11}$ both above and below $T_N$ confirming the
existence of important magnetic fluctuations in this
geometrically frustrated magnet, fluctuations that are likely
enhanced by a magnetic field. It has been suggested from heat
capacity data \cite{TACHIBANA2005} that a single Einstein
contribution could adequately fit the low-temperature deviation,
indicating that there is no evidence of a anomalous magnetic
contribution for $T < T_N$. Our data do not support this
suggestion.

\section{CONCLUSION}
In conclusion, our ultrasonic experiments have identified an
important coupling between the AF magnetic order parameter and
the lattice in hexagonal YMnO$_3$. This coupling yield important
stiffening anomalies below $T_N$ only for in-plane elastic
deformations, compression and shear. This implies that the
coupling is established via the modulation of the in-plane
exchange interactions. The critical exponent $\beta$ for the AF
order parameter near $T_N$ is found higher than the expected value
from XY and 3D Heisenberg universality classes. This situation
favors a conventional antiferromagnetic long range ordering as
suggested from other measurements. However, the data confirm
important spin fluctuations over a wide temperature range above
and below the N\'eel temperature. Finally, a sharp step appearing
on $C_{11}$ at $T_N$, for which hysteresis is observed, could be
consistent with a weak coupling between ferroelectricity and
magnetism at domain walls.

\acknowledgments{The authors thank G. Quirion, S. Jandl and C.
Bourbonnais for discussions and M. Castonguay for technical
support. This work was supported by grants from the Fonds
Qu\'eb\'ecois de la Recherche sur la Nature et les Technologies
(FQRNT) and from the Natural Science and Engineering Research
Council of Canada (NSERC).}

\end{document}